\documentstyle[preprint,aps]{revtex}
\begin{document}
\title{ Spacetime Duality\\
 and Two--dimensional Gauge Field Theory }
\author{A.Smailagic \footnote{E-mail 
address: anais@drava.etfos.hr}}
\address{Faculty of Electrical  Engineering \break
University of Osijek, Croatia}

\author{Euro Spallucci\footnote{E-mail 
address: spallucci@trieste.infn.it}}
\address{Dipartimento di Fisica Teorica\break
Universit\`a di Trieste,\\
INFN, Sezione di Trieste}
\maketitle

\begin{abstract}
In this letter we implement a recently proposed {\it spacetime duality}
approach to dualize a  two dimensional, Abelian, gauge field theory,
which has no dual version under $p$--duality. Our result suggests that
spacetime duality spans a new, wider, class of dual theories, which cannot
be related one to another by  $p$--duality transformations.
\end{abstract}

\newpage
\noindent
	Recently, a new proposal for dualizing field theories 
	has been  put forward \cite{gris}. Contrary to the 
	approach based on internal gauge symmetries, known as
	p--duality (a good review of   p--dualities, with an
	extensive reference list can be found in \cite{pdual} ),
	the new approach exploits local spacetime  symmetry as the basic 
	tool to  obtain a dual theory. In \cite{gris} such approach  has been 
	applied in two dimensions to scalar and fermion theories,
	as well as to their super--symmetric extensions. The resulting dual
	model was  a (super) scalar field theory. Only in the case  of $(2,2)$ 
	super--symmetric models,  dual theory exhibits a certain degree of 
	complexity due to the exchange of  chiral and twisted chiral
	super fields.
	These results are in agreement with the p--duality approach  relating 
	$p$--forms to  $D-p-2$ dual forms. Accordingly, for $D=2$ the only
	allowed duality relation is between $p=0$ forms, i.e. 
	scalar fields are dual to scalar fields.
	Nonetheless, spacetime approach to dualities bears no
	obvious requirement of equivalence to $p$--dualities, because 
	the two approaches follow from  different symmetries.\\
	
	\medskip
	\noindent
	In this letter we would like to show that the spacetime approach allows
	to establish new duality relations which cannot be  accounted for 
	in the p--duality approach.  The massless, Abelian, vector model,
	in two dimensions, is a gauge theory without a dual version in the 
	p--duality sense, but can be consistently coupled to gravity and
	dualized following the path integral formulation outlined 
	in \cite{gris}. The classical action for the gauge field $A_\mu$ coupled 
	to the background metric $h_{\mu\nu}$  is given by	
	
	\begin{equation}
	S[\,\,h\, ,A\, , {\overline c}\, , c\,\,]= 
	\int d^2x \sqrt{-h}\left[\,-{1\over 4}
	h^{\mu\alpha}\, h^{\nu\beta}\, F_{\mu\nu}\, F_{\alpha\beta}-{1\over 2}
	\left(\,\nabla_h{}^\mu A_\mu\, \right)^2 + {\overline c} \, {\bf \Box}_h \,
	 c\,\right]\label{uno}
	\end {equation}
	where, $\nabla_h{}^\mu $ is the covariant derivative with respect to the 
	background metric and ${\bf \Box}_h $ is the covariant D'Alembertian.
	We have also introduced a gauge fixing term and a pair
	of Fadeev--Popov ghosts ${\overline c}$, $c$. Having coupled the gauge
	field to a background metric we can define a non--trivial generating 
	functional as 
	
	\begin{equation}
	Z[\,h\,]=\int [\,{\cal D}A\,]_h\, [\,J_{FP}\,]_{A,h}\,
	\exp iS[\,h,A\,]\label{due}
	\end{equation}
	In (\ref{due}), the ghost action has been written in terms of
	the  Fadeev--Popov determinant $[\,J_{FP}\,]_{A,h}$.\\
	The basic difficulty in functional integration is  the definition
	of  the functional measure. Quantum anomalies can appear
	as a result of the impossibility to define a functional measure
	preserving all the classical symmetries \cite{fuji}.
	If one insists on having a  covariant measure necessarily ends up
	producing a conformal anomaly. 
	Therefore, it is useful to put in evidence conformal
	degree of freedom within the functional measure.  For this purpose
	one first chooses a decomposition of the metric as 
	$\displaystyle{h_{\mu\nu}=\hat h_{\mu\nu}\, e^\varphi}$, where
	  $\hat h_{\mu\nu}$ is the traceless part of $ h_{\mu\nu}$. Then,
	the measure of an arbitrary field $X$  transforms as  
	
	\begin{equation}
	[\,{\cal D}X\,]_h=  [\,{\cal D}X\,]_{\hat h}
	\, J[\,\hat h\, ,\varphi\,]\label{tre}
	\end{equation}
	where, $ J[\,\hat h\, ,\varphi\,]$ is the Jacobian corresponding to the 
	above decomposition of the metric. The Jacobian is a local quantity 
	and can be written in terms of the local part of the anomalous
	gravitational action \cite{ddk} given by the Liouville action
	$S_L^X[\,\hat h,\varphi\,]$ :
	
	\begin{equation}
	S_L^X[\,\hat h,\varphi\,]= \int d^2x\sqrt{-\hat h}\left[\,{1\over 2}
	\varphi\, {\bf \Box}_ {\hat h} \,\varphi + R_h\,\varphi   \,\right]
	\label{liouv}
	\end{equation}
	The generating functional (\ref{due}) can then be written as
	
	\begin{equation}
	Z[\,h\,]=\int [\,{\cal D}A\,]_{\hat h}\, J[\,\hat h\, ,\varphi\,]\,
	 [\,J_{FP}\,]_{A,\hat h}\,
	\exp iS[\,\hat h\, ,\varphi\, , A\,]\label{genfunct}
	\end{equation}
	One may notice appearance of the gauge Fadeev--Popov
	determinant in the
	form $[\,J_{FP}\,]_{A,\widehat h}$ instead of the originally introduced
	$[\,J_{FP}\,]_{A,h}$, which is due to the  conformal invariance of the
	 scalar ghost action for the gauge field.\\
	Gravitational anomalous action can be obtained from (\ref{genfunct})
	integrating out the gauge field.\\
	Classical actions for massless scalar and fermion field are conformally 
	invariant in two dimensions, while the gauge field action is not. 
	Classical non--invariance has to be appropriately accounted for
	within the path integral approach. \\
	Any classically non--invariant action can be split  
	into an invariant piece and a classical trace contribution 
	$\displaystyle{S^{\, non-inv}[X,h]=S [X,\hat h]+S^{\, tr}[X, \hat h, \varphi]}$. 
	The invariant piece of the classical action  contributes a numerical 
	coefficient $k_X$ to the anomalous gravitational action, while the trace 
	part contributes a different numerical coefficient $k^{\, tr}_X$. Both
	coefficients can be computed  from the {\it finite} part of the 
	anomalous Feynman graph a la' Adler--Rosenberg
	\cite{gris2}, in diagrammatic perturbation theory, or can be extracted 
	from the finite part of the regularized Jacobian in the Fujikawa 
	approach \cite{fuji}. In addition there is a {\it finite and local}
	contribution to the gravitational anomalous action which is extracted,
	 in the process of regularization,  from the
	{\it divergent} part of the anomalous diagram and
	 whose numerical coefficient is  an arbitrary, regularization dependent,
	parameter ``$a$'' \cite{noi}.  The freedom in the choice of  ``$a$'' 
	allows one to shift among various regularization schemes. 
	This coefficient is now appearing in front of the Liouville action
	of the functional measure and
	is  the same both for classically invariant  and for classically 
        non--invariant theories. The evaluation of the functional 
        integral for classically non--invariant theories proceeds as follows

	\begin{eqnarray}
	Z[\,h\,]&=&\int [\,DX\,]_h \exp\left(iS[\,X\, ,h\,]\right)\nonumber\\
	&=&\exp \left(iaS_L[\,\varphi\, ,\hat h\,]\right)\int [\,DX\,]_{\hat h}
	\exp\left(iS [\,X\, ,\hat h\,]+iS^{\, tr}[\, X\, ,\hat h,\varphi\,]
	\right)\nonumber\\
	&=&\exp \left(iaS_L[\,\varphi\, ,\hat h\,]+
	i k_X S[\, \hat h\, ]+ik_X^{\, tr} S[\, \hat h\, , \varphi\,]\right)\label{non}
	\end{eqnarray}
	where, 
	$\displaystyle{S[\, h\,]=\int d^2 x\sqrt{-h}\, R(h)\,\Box_h^{-1}\, R(h)}$ 
	 and various numerical coefficients have been put in evidence.
	The above result generalizes the path integral anomaly calculation to
	the case of  classically conformally non--invariant actions. In order 
	to compare (\ref{non}) to the result of an explicit calculation,
	 such as in \cite{gris2}, one has to choose a definite regularization 
	scheme. Suppose one wants to define a diffeomorfism invariant measure. 
	This requires
	$\displaystyle{a\,S_L[\,\varphi\, ,\hat h\,]+ k_X\, S[\,\hat h\,]= k_X\,(
	S_L[\,\varphi\, ,\hat h\, ]+S[\,\hat h\,])= k_X\, S[\, h\, ]
	}$
	which gives $\displaystyle{a=k_X}$ within a covariant regularization 
	scheme. One can give a general formula for evaluating numerical
	coefficients of the invariant piece of the
	anomalous gravitational action through the relation \cite{noi2}
	\begin{equation}
	\left[ det\, D_q\,\right]^{(-N/2)^{s+1}}=
	\exp\left[i {N\over 48\pi}(-1)^s \left(6q^2 -6q + 1\right)S[g]
	\right]\label{numeri}
	\end{equation}
	where, $q$ is the Lorenz weight of the field, $  D_q$ is the covariant
	operator acting on that field, $N$ is the number of fields and $s$
	 denotes statics of the field ($s=0$ for bosons, $s=1$ for fermions).
	From (\ref{numeri}) one sees that $k_X= N(-1)^s (6q^2
	-6q + 1)$.   \\
	The anomalous gravitational action for a classically conformally
	non--invariant action turns out to be 
	
	\begin{equation}
	Z[\,h\,]=\exp\left(ik_X^{\, non-inv}S[\, h\, ]\right)
	\end{equation}
	with $\displaystyle{k_X^{\, non-inv}=k_X+k_X^{\, tr}}$.  
	The  path integral evaluation of gravitational
	anomalies   thus reproduces
	results obtained in \cite{gris2} by other methods. 
	Classically conformally invariant action  result is obtained  simply 
	putting $k_X^{\, tr}=0$. With the above generalization of functional 
	integration of matter fields in mind, one proceeds on the path of 
	dualizing gauge theory. 
	    Following \cite{gris}, we promote the background metric
	$h_{\mu\nu}$ into a dynamical gravitational field $g_{\mu\nu}$ which
	serves to dualize  the original action (\ref{uno}).
	The corresponding {\it mother } generating functional is defined as
	
	\begin{equation}
	Z[\,h\,]=\int [\,{\cal D}A\,]_g\, [\,{\cal D}g\,]_g\,
	 \Delta[\,g,h\,]\, [\,J_{FP}\,]_{A,g}\, \exp iS[\,g,A\,]
	\label{duebis}
	\end {equation}
	$\Delta[\,g,h\,]$ denotes a constraint condition  to be chosen
	 on the gravitational field   $g_{\mu\nu}$
	in such a way to restrict it  to the background
	field $h_{\mu\nu}$. Then, (\ref{duebis}) will
	reproduce the starting functional (\ref{due}). 
	One chooses the ``~conformal gauge~'' condition
	$\displaystyle{g_{\mu\nu}=e^\phi\, h_{\mu\nu} }$.
	This gauge choice fixes
	the diffeomorfism transformations and, thus, introduces in (\ref{duebis})
	a gravitational Fadeev--Popov determinant which can be expressed in 
	terms of the gravitational ghost anomalous action as 
	 $\displaystyle{[\,J_{FP}\,]_{g}=\exp
	 \left(\, i k_{\, gh}\, S[\, g\,]\right)}$ where $k_{\, gh}$ is the ghost
	 coefficient. As an example of calculation of the numerical
	coefficient,  take a pair 
	of gravitational ghost--antighost ($B_{\mu\nu}$, $C_\mu$) corresponding
	to the Fadeev--Popov determinant of the conformal gauge for which
	$q_B=2$, $N=2$, $s=1$. Then, one finds the known value $k_{\, gh}= -26$.\\
	One  defines the constraint $\Delta[\,g,h\,]$
	as a functional integral over the Lagrange multiplier
	field $\Lambda$ as
	
	 \begin{equation}
	\Delta[\,g,h\,]=\int [\,{\cal D}\Lambda\,]_g \,
	G(g)\exp\left[\, -i\int d^2x\sqrt{- g}\,
	\Lambda \, F(h,\phi)\,\right]\label{lag}
	\end{equation}
	where $	G(g)$ is to be determined in the process of reproducing the
	original matter generating functional (\ref{due}). The functional
	integral (\ref{lag}) is introduced in order
	to remove conformal degrees of freedom. Therefore, it should be 
	proportional to the delta function $\delta[\,\phi\,]$. 
	 This imposes that the integrand in the exponential factor be linear
	 in the conformal degree of freedom while the rest must depend on the
	 background metric only as
	$\displaystyle{\sqrt{- g}\, F(h,\phi)=\sqrt{-h}\, F_h\, \phi}$.  The 
	{\it mother functional,} can now be written,  as
	
	\begin{eqnarray}
	Z[\,h\,]=&&\int [\,{\cal D}A\,]_h \,  [\,{\cal D}\phi\,]_h  \,
	   [\,{\cal D}\Lambda\,]_h\,
	[\,J_{FP}\,]_{A,g}\,    [\,J_{FP}\,]_{g}\, G(g)\,
	\exp iS[\,g,A\,]\times\nonumber\\ 
	&&\exp i\left[\,k_A S_L{}^A[\,g\,] + k_\phi S_L{}^\phi[\,g\,] +k_\Lambda
	 S_L{}^\Lambda[\,g\,]\,\right]
	{}\exp\left[\, -i\int d^2x \sqrt{-h}\,\Lambda\, F_h\phi \,\right]
	  \label{quattro}
	\end{eqnarray}
	where, we have put in evidence the conformal degree of freedom 
	of the functional measure through  the Liouville actions
	with corresponding numerical coefficients.  The 
	notation bears explicit reference to different fields 
	in order that the reader can keep track of various terms 
	coming from different functional measures. It may
	be worth to remark that all the Liouville actions have the same form
	(\ref{liouv}) and could be re--written as 
	$\left(\,\sum_f k_f\,\right)S_L[\,g\,]$, where $\sum_f$ means summing over
	various fields.\\
	Basic requirement on the  mother functional is to
	 reproduce the starting, gauge, generating functional after integrating
	 out all the other field variables but $A_\mu$. Let us start with the  
	 integration of the Lagrange multiplier. One obtains
	
	\begin{equation}
	\int [\,{\cal D}\Lambda\,]_h\,\exp\left[\, -i\int d^2x \sqrt{-h}\,
	\Lambda\, F_h\phi\,\right]=\delta[\,F_h\phi\,]= 
	\left(\, det\; F_h\, \right)^{-1}\, \delta[\,\phi\,]
	\end{equation}
	which enables us to  rewrite the mother functional as 
	\begin{eqnarray}
	Z[\,h\,]=&&\int [\,{\cal D}A\,]_h\,[\,J_{FP}\,]_{A,h}\, \exp iS[\,h,A\,]  
	[\,{\cal D}\phi\,]_h\, \delta[\,\phi\,]\,
	\left(\, det\; F_h \, \right)^{-1}\times\nonumber\\
	     && [\,J_{FP}\,]_{g}\, G(g)\,
	\exp \left[\,i\, (\sum_f k_f) S_L[\, g\,] \,\right]
	\exp iS^{\, tr}[\, h,\phi,A\,] \label{cinque}
	\end{eqnarray}
	
	Integration over the delta function removes $\phi$--dependent terms
	giving
	
	\begin{equation}
	Z[\,h\,]=\int [\,{\cal D}A\,]_h\,[\,J_{FP}\,]_{A,h}\, \exp iS[\,h,A\,]
	 \left(\, det\; F_h \, \right)^{-1}[\,J_{FP}\,]_{h}\, G(h)\,
	 \label{cinquebis}
	\end{equation}
	 From (\ref{cinquebis}) one can see that the starting functional
	 (\ref{due}) can be reproduced if 
	 
	 \begin{equation}
	\left(\, det\; F_h \, \right)^{-1}[\,J_{FP}\,]_{h}\,
	G(h)=1\label{one}
	\end{equation}
	 The above condition determines the form $G(h)$ in terms of
	 the Fadeev--Popov determinant for gravitational ghosts and the
	 determinant of the operator $F_h$. This, however, does not determine
	 the part of $G(h,\phi)$ which has been eaten by the delta function
	 $\delta(\phi)$. We eliminate this arbitrariness by the choice 
	 $  G(h,\phi) \,[\,J_{FP}\,]_{h,\phi}=1 $.
	 Inserting (\ref{one}) in (\ref{quattro}) one can write a more
	convenient form of the mother generating functional which serves to
	produce the dual theory as
	 \begin{eqnarray}
	Z[\,h\,]=&&\int [\,{\cal D}A\,]_h\, [\,{\cal D}\phi\,]_h\,[\,{\cal
	D}\Lambda\,]_h\,[\,J_{FP}\,]_{A,h}\, \exp iS[\,g,A\,]  
	\,\left(\, det\; F_h\, \right)\times\nonumber\\
	&&\exp\left[\, i\,(\,\sum_f k_f\,)\, S_L[\,g\,] \,\right]\,
	\exp\left[\, -i\int d^2x \sqrt{-h}\,\Lambda\, F_h\phi \,\right]
	\label{dual}
	\end{eqnarray}
	From this point on, we  shall  follow the dual route. 
	Starting with (\ref{dual}) one performs a reversed order of functional
	integration and first integrate out the gauge field to obtain:
	
	\begin{eqnarray}
	Z^{\, dual}[\,h\,]=\left(\, det\; F_h\, \right)\,\exp \left(
	ik_A^{\, non-inv}S_A [\,h\,]\,\right)
	\int [\,{\cal D}\Lambda\,]_h\, [\,{\cal D}\phi\,]_h\, 
	&&\exp \left[i\,(\,\sum_{f^{\,\prime}} k_{f^{\,\prime}}\,)\, S_L[\,g\,] \,\right]
	\times\nonumber\\
	&&\exp\left[\, -i\int d^2x \sqrt{-h}\,\Lambda\, F_h\phi \,\right]
	\end{eqnarray}
	where, $\sum_{f^{\,\prime}}=\sum_f + k_A^{\, tr} $.
	The explicit form of the Liouville action 
	(\ref{liouv})  can be written in a condensed notation as~\footnote{
	We have rescaled conformal degree of freedom in such a way
	to absorb numerical coefficient $\sum_{f^{\,\prime}} k_{f^{\,\prime}}$. }
	 $\displaystyle{S_L[\,h,\phi\,]=
	\int d^2x \sqrt{-h}\,\left(\,\phi\, L_h\, \phi + 
	M_h\, \phi\,\, \right)}$ , where $L_h$ and $M_h$ stand, respectively,
	 for the Liouville kinetic operator and the  
	 non--minimal, linear  coupling   between $h_{\mu\nu}$ and
	 $\phi$. The Gaussian integration over $\phi$ gives

	\begin{eqnarray}
	Z^{\, dual}[\,h\,]=&&\left(\, det\; F_h\, \right)\, 
	\exp\left( ik_A^{\, non-inv}S _A [\,h\,]\,\right)
	\,\int [\,{\cal D}\Lambda\,]_h\,
	 \left(\, det\; L_h\, \right)^{-1/2}\times\nonumber\\
	&&\exp\left[\, -i\int d^2x \sqrt{-h}
	\left(\,\,\Lambda\, F_h + M_h \,\, \right)
	\,L_h^{-1}\, \left(\,\,\Lambda\, F_h + M_h \,\, \right) \right]
	\end{eqnarray}
	To  simplify the above expression we introduce the new field variable
	
	\begin{equation}
	\hat\phi= L_h^{-1}\, \left(\,\,\Lambda\, F_h + M_h \,\, \right)
	\end{equation}
	Accordingly, the functional measure for the Lagrange multiplier field
	transforms as
	
	\begin{equation}
	[\,{\cal D}\hat\phi\,]_h =\left(\, det\; L_h\, \right)^{-1}\,
	\left(\, det\; F_h\, \right)\, [\,{\cal D}\Lambda\,]_h
	\end{equation}
	and the  dual generating functional turns out to be of the form
	
	\begin{equation}
	Z^{\, dual}[\,h\,]=\left(\, det\; L_h\, \right)^{1/2}\,
	\exp \left(ik_A^{\, non-inv}S _A[\,h\,]\right)\, 
	\int[\,{\cal D}\hat\phi\,]_h\,
	\exp\left[ -i\int d^2x \sqrt{-h}\,\hat\phi\, L_h\,
	\hat\phi\,\right]
	\label{final}
	\end{equation}
	From (\ref{final}) one can see the cancellation of $\left(\, det\; F_h\,
	\right)$ which shows that the actual form of $F_h$ is not
	really important. One could have as well started with the simple
	$\delta(\phi)$ corresponding to the choice $F_h=1$. The only reason
	we can think  in favor of choosing a covariant constraint 
	$\delta\left(\, R(g)-R(h)\,
	\right)\equiv\delta\left(\,\Box\,\phi\,\right)$,
	exploited in \cite{gris}, is purely aesthetical, and because it also mimics
	covariant Lorenz gauge fixing in $p$--duality approach.  
	The coefficient $ k_A^{\, non-inv} $ has been calculated in \cite{gris2} 
	and results to vanish, $ k_A^{\, non-inv}=0$. Therefore, (\ref{final}) 
	reduces to

	\begin{equation}
	Z^{\, dual}[\,h\,]=\left(\, det\; L_h\, \right)^{1/2}\,
	\int[\,{\cal D}\hat\phi\,]_h\,
	\exp\left[ -i\int d^2x \sqrt{-h}\,\hat\phi\, L_h\,
	\hat\phi\,\right]
	\end{equation}
	The dual action (\ref{final})turns out to be of  the same form for all
	fields with
	numerical coefficients of the conformally invariant piece of the
	classical action given by (\ref{numeri}), while the trace coefficients
	(if any) have to be calculated case by case. 
	 The scalar (fermion) anomalous action cancels against
	 $\left(\, det\; L_h\, \right)$. The final result is a 
	scalar--to--scalar duality, exactly like in the p--duality
	approach, or fermion bosonization \cite{gris}.
	In the case of the gauge field, there is no cancellation, and we are 
	left with the determinant of the Liouville kinetic operator 
	$L_h$ together with the  scalar field $\hat\phi$.   
	One can re--write the remaining determinant as a functional integral of
	{\it another} scalar field which, however, has to be ghost--like due
	 to the positive power of the determinant exponential factor $+1/2$. 
	 So, we obtain the action dual to the gauge field action as
	\begin{equation}
	Z^{\, dual}[\,h\,]=
	\int[\,{\cal D}\hat\phi\,]_h  \,  [\,{\cal D}\xi\,]_h\, 
	\exp\left[ -i\int d^2x \sqrt{-h}\left(\,\,
	\hat\phi\, L_h\,\hat\phi +
	\xi\, L_h\,\xi\,\, \right)
	\right]\label{finale}
	\end{equation}
	
	\medskip
	\noindent
	We have  shown that dual action for the gauge field in two dimensions 
	exist within the spacetime approach to dualities. The result 
	(\ref{finale}) cannot be obtained in the p--duality approach,
	 and suggests the existence of a new, larger, class of dual theories. 
	With hindsight, a non--trivial  dual action for a two dimensional
	vector gauge field is not a surprise and can be explained as follows.
	In the widest sense, dual theories
	should be such to preserve the number of degrees of freedom in
	passing from one to another. The explanation of our result is then 
	clear. One starts with the gauge field carrying no physical degrees 
	of freedom in two dimensions, and ends up with the dual theory written 
	in terms of one physical scalar field and one scalar ghost which, together,
	 still carry no physical degrees of freedom. The price one has to pay 
	 is that the dual theory contains more than one field, which is not 
	 high price at all. The possibility to have 
	 a wider class of dual theories in the above sense is by no means
	inherent to two dimensions. The procedure described in this
	letter has no obvious dependence on the dimensionality of the physical
	spacetime. It should be, therefore, applicable to higher than two
	dimensions provided the Liouville action remains local and Gaussian. 
	The locality of the Liouville action is expected on the grounds
	that it is a synonym for a regularization  
	counter--term contribution, as explained earlier, which is always local. 
	In particular, the four dimensional analogue of the two--dimensional
	gauge field is the third rank anti--symmetric tensor field 
	with no dynamical degrees of freedom.
	It is expected, therefore, to  be dualized in terms of 
	scalar and ghost--like fields in the above sense.  
	Moreover, there is no {\it a priori} reason for higher dimensional
	cancellation of the various determinants in   (\ref{final}), even for 
	scalar fields.  It is reasonable to expect that dual 
	theories will in general belong to the wider class of dualities 
	described in this paper. We hope to return to this point in future.

	\end{document}